\documentclass[pra,twocolumn,amsmath]{revtex4}
\begin{document}
\def\ket#1{|#1\rangle} 
\def\bra#1{\langle#1|}
\def\av#1{\langle#1\rangle}
\def\myarrow{\mathop{\longrightarrow}}

\title{Constraints for quantum logic arising from conservation laws and field fluctuations}

\author{Julio Gea-Banacloche}
\email[]{jgeabana@uark.edu}
\affiliation{Department of Physics, University of Arkansas, Fayetteville, AR 72701, USA}
\author{Masanao Ozawa}
\email[]{ozawa@math.is.tohoku.ac.jp}
\affiliation{Graduate School of Information Sciences, Tohoku University, Aoba-ku, Sendai 980-8579, Japan}

\date{\today}

\begin{abstract}
We explore the connections between the constraints on the precision of quantum logical operations that arise from a conservation law, and those arising from quantum field fluctuations.  We show that the conservation-law based constraints apply in a number of situations of experimental interest, such as Raman excitations, and atoms in free space interacting with the multimode vacuum.  We also show that for these systems, and for states with a sufficiently large photon number, the conservation-law based constraint represents an ultimate limit closely related to the fluctuations in the quantum field phase. 
\end{abstract}
\maketitle

\section{Introduction.}

It was pointed out in \cite{jgb1} that, when trying to do quantum logic with a quantized field, three possible kinds of errors could arise:  errors due to qubit-field entanglement (a point also made in \cite{vanenk}), to field phase fluctuations, and to field amplitude fluctuations.  The last two are easy to separate from each other, but it is not immediately clear how they relate to the first.

Almost simultaneously, and quite independently, it was also pointed out \cite{ozawa1} that the existence of a conservation law, which holds for many of these quantized-field systems, would also constraint the accuracy achievable in some quantum logical gates.  It is natural to ask the question of how this result may relate to the ones derived in \cite{jgb1} by very different methods, and this is, broadly speaking, the purpose of the present paper.  

The remainder of this introductory section is devoted to some preliminary observations, followed by a brief summary of the rest of the paper's sections and results.

It seems that the conservation-law induced quantum limit (which we shall abbreviate as CQL from now on, for consistency with \cite{ozawa1b}) is most directly related to the entanglement error:  if the conservation law is of the form
\begin{equation}
(\pm)\frac{1}{2} \sigma_z + a^\dagger a = \,\hbox{const} 
\label{one}
\end{equation}
then it is impossible, starting from a factorized state that is an eigenstate of $\sigma_z$, to produce a {\it pure\/} eigenstate of $\sigma_x$---that is, to accomplish a transformation like $\ket 0 \to (\ket 0 + \ket 1)/\sqrt 2$ without getting entangled with the field.  Intuitively, it is clear that if the transition from $\ket 0$ to $\ket 1$ requires the absorption of a photon, then the state of the field associated with the state $\ket 1$ in the above superposition will, in general, be different from the state associated with $\ket 0$.  Nonetheless, the actual amount of entanglement depends on the field state and the coupling assumed, and the question has proved to be somewhat subtle (in particular, for coherent field states, which are not changed by the loss of one photon).  We refer the readers to the work of Silberfarb and Deutsch \cite{silberfarb} on this question, and also to the papers \cite{itano,vanenk2,jgb2} for other related aspects of the controversy. (We will comment later on another issue that was clarified in \cite{itano,jgb2}, namely, the relationship between the constraints in \cite{jgb1} and those posed by spontaneous emission for an atom in free space.)  

There is still the question of how this relates to phase and/or amplitude fluctuations.  The form of the CQL is (in a way that we shall make precise later; see Section II)
\begin{equation}
P_e \ge \frac{1}{4}\,\frac{1}{1 + 4 \sigma(n)^2}
\label{two}
\end{equation}
where $\sigma(n)$ is the standard deviation of the number operator $n\equiv a^\dagger a$ in the initial state of the field.  $P_e$ is the ``error probability'' for the quantum logical operation, defined as the largest deviation (in the norm-square sense) between the intended final state of the qubit and the actual final state, maximized over all input qubit states.   Eq.~(\ref{two}) indicates that this error probability {\it decreases\/} as the fluctuations in the field intensity {\it increase\/}.  Thus, it seems clear that this error is {\it not\/} due to field-intensity (or, probably, field-amplitude) fluctuations.  There is, however, a suggestion that (at least in some limit) it might be related to {\it phase\/} fluctuations, since for these a relationship of the form  $\sigma(\phi)^2 \ge 1/4\sigma(n)^2$ is generally understood to hold for some reasonable approximate Hermitian phase operator $\phi$ (see, for instance, \cite{barnett}, and also the discussion below, in Section IV).

The suggestion is, therefore, that the CQL (\ref{two}) may be related to the error caused by fluctuations in the field quantity conjugate to the one that appears in the conservation law (\ref{one}). This is explored, and a precise correspondence is established, in Section IV of this paper.  Prior to this, in Section II, we present a simple proof of the result (\ref{two}), and in Section III we show some examples of systems of interest where a conservation-law-induced constraint applies.  Of particular interest are a Raman-coupled system (where the conserved quantity is less obviously related to the total angular momentum), and an atom in free space, where spontaneous emission is the principal limiting factor. Section V briefly summarizes our conclusions. 

\section{Derivation of the CQL.}

This section presents a simplified derivation of the constraint on quantum logic  arising from a conservation law, specialized for our present purposes; see \cite{ozawa2} for a much more detailed derivation and additional context.

In what follows we assume that we are trying to perform a quantum logical operation (typically a Hadamard gate) on a qubit using a Hamiltonian that involves the qubit as well as a quantized control field, subject to a conservation law of the form $L=L_1+L_2=$ const, with $L_1$ a qubit operator and $L_2$ a field operator.  Typically $L_1$ will be the Pauli operator $\sigma_z$, written in the computational basis $\{\ket 0,\ket 1\}$ as
\begin{equation}
\sigma_z = \ket 0 \bra 0 - \ket 1 \bra 1
\label{three}
\end{equation}
The identification of the states $\ket 0$ and $\ket 1$ with appropriate states of the qubit (typically an atom) will depend on the system (see next Section for examples).  

We therefore assume that the field-qubit interaction Hamiltonian is such that after a certain time $t$ the evolution operator $U(t)$ (the time argument will be omitted below for simplicity, as long as no ambiguity results) \emph{approximately} carries out a Hadamard transformation, while leaving invariant the quantity
\begin{equation}
L = \sigma_z + L_2
\label{four}
\end{equation}
We note that the exact Hadamard transform goes as follows:
\begin{align}
\ket 0 &\to \frac{1}{\sqrt 2}\left(\ket 0 + \ket 1\right) \notag \\
\ket 1 &\to \frac{1}{\sqrt 2}\left(\ket 0 - \ket 1\right) 
\label{five}
\end{align}
and therefore it changes the operator $\sigma_z$ into $\sigma_x = \ket 0\bra 1 + \ket 1\bra 0$.  Accordingly, we may take as a measure of the success of the gate the smallness of the difference (or ``error'') operator $D$ defined as
\begin{equation}
D = U^\dagger \sigma_z U - \sigma_x
\label{six}
\end{equation}
Consider now the commutator $[\sigma_x+D,L] = [U^\dagger\sigma_zU,L]$. We have
\begin{align}
[\sigma_x+D,L] &= U^\dagger \sigma_z U L - L U^\dagger \sigma_z U \notag \\
&= U^\dagger \sigma_z U U^\dagger L U - U^\dagger L U  U^\dagger \sigma_z U \notag \\
&= U^\dagger[\sigma_z,L]U\notag \\
&=0
\label{seven}
\end{align}
since, by assumption, $U^\dagger L U = L$, and $\sigma_z$ commutes with $L$ (it clearly commutes with itself, and it also commutes with $L_2$, at equal times, because they act on different systems).  From this we conclude
\begin{equation}
[L,D] = [\sigma_x,L] = -2i\sigma_y
\label{eight}
\end{equation}
and therefore the standard deviations of $D$ and $L$ in an arbitrary (initial) state of the joint qubit-field system must satisfy
\begin{equation}
\sigma(D)\sigma(L) \ge \frac{1}{2}\left|\av{[\sigma_x,L]}\right| = |\av{\sigma_y}|
\label{nine}
\end{equation}
The right-hand side has its maximum value (equal to 1) in an eigenstate of $\sigma_y$, so we find
\begin{equation}
\sigma(D)_{max}^2 \ge \frac{1}{\sigma(L)^2} = \frac{1}{1+\sigma(L_2)^2} 
\label{ten}
\end{equation}
where the last equality follows if one assumes that the initial state of field and qubit is a product state (note $\sigma(\sigma_z)^2 = 1$ in an eigenstate of $\sigma_y$).  

Equation (\ref{ten}) is the basic constraint on the accuracy of a Hadamard gate in the presence of a conserved quantity of the form (\ref{four}).  In the remainder of the paper we shall work mostly with this form, but it is useful to relate the error measure $\sigma(D)_{max}^2$ to the more familiar ``gate fidelity'' $F$ that is defined in terms of the overlap of the final qubit state with the intended state.  As $1-F^2$ gives the probability to find the qubit in a state orthogonal to the intended one, this quantity may be termed the failure (or ``error'') probability of the gate, $P_e$.  Taking $F$ to be the ``worst case'' fidelity, over the set of all possible initial states of the qubit, it can be shown (see Appendix A below), along the lines of \cite{ozawa2}, that 
\begin{equation}
P_e = 1 - F^2 \ge \frac{1}{4}\av{D^2}_{max}\ge \frac{1}{4}\sigma(D)^2_{max} \ge  \frac{1}{4}\, \frac{1}{1+\sigma(L_2)^2} 
\label{eleven}
\end{equation}
As an example, for a system for which the quantity (\ref{one}) (proportional to the total angular momentum of the system) is conserved, we would write $L = \sigma_z \pm 2 a^\dagger a$, so $L_2 = 2a^\dagger a$, and then (\ref{eleven}) would immediately yield the result (\ref{two}).  We discuss this further, and provide several additional examples, in the following Section. 

Although the discussion above has been very specific, it is straightforward to see how all these arguments can be generalized to arbitrary systems and operations.  The key ingredients are: (1) a conserved quantity $L$ that can be written as the sum of a qubit operator $L_1$ and a control operator $L_2$, and (2) a desired evolution that would change $L_1$ into another qubit operator that does not commute with $L_1(0)$.  One then finds that (3) such an operation cannot, in general, be carried out exactly, and the unavoidable error (for at least some initial states) is inversely proportional to the sum of the uncertainty squared of $L_1$ and $L_2$ in the initial state.

\section{Examples of systems where the CQL applies.}

\subsection{Two-level atom and single-mode field (Jaynes-Cummings model).}

The simplest system where a conserved quantity of the form (\ref{one}) is found to hold is the well-known Jaynes-Cummings model \cite{jcm}, which describes the coupling of a two-level atom with a single-mode of the electromagnetic field.  Allowing for a detuning $\Delta$ between the atom and the field, the Hamiltonian for the system may be written in a suitable interaction picture as
\begin{equation}
H = \hbar\Delta a^\dagger a + i\hbar g \left(a\otimes\ket e \bra g - a^\dagger \otimes \ket g \bra e\right)
\label{twelve}
\end{equation}
where $g$ is an appropriate coupling constant.  To see how this Hamiltonian performs an approximate Hadamard gate, replace the field operators $a$ and $a^\dagger$ by classical constants $E_0 e^{i\phi}$ and $E_0 e^{-i\phi}$, respectively, and make the identification $\ket e \to \ket 1$, $\ket g \to \ket 0$.  Then the evolution operator, for zero detuning, is 
\begin{equation}
U(t) = \cos(g E_0 t) + \begin{pmatrix} 0 & -e^{-i\phi} \\ e^{i\phi} & 0 \end{pmatrix} \sin(g E_0 t)
\label{thirteen}
\end{equation}
in the $\{\ket 0,\ket 1\}$ basis, and at the time $t=\pi/4gE_0$, if $\phi=0$, this becomes
\begin{equation}
U=\frac{1}{\sqrt 2}\begin{pmatrix} 1 & -1 \\ 1 & 1 \end{pmatrix}
\label{fourteen} 
\end{equation}
This is not actually a Hadamard gate $H$, but rather $\sigma_x H$ (the state $\ket 1$ is changed to $\ket 1 -\ket 0$).  Nonetheless, it is easily verified that the operation $U^\dagger \sigma_z U$ produces $-\sigma_x$, and hence the results in the previous section apply if the difference operator $D$ is defined as $D=U^\dagger \sigma_z U+\sigma_x$ and the sign of $\sigma_x$ is changed in the equations that follow (\ref{six}).

It is trivial to verify that the quantity $-\frac{1}{2}\sigma_z + a^\dagger a$ is conserved under the Hamiltonian (\ref{twelve}) (regardless of whether the detuning $\Delta$ is zero or not), and therefore Eq.~(\ref{ten}) holds, with $L_2 = -2 a^\dagger a$.  By Eq.~(\ref{eleven}), this yields a failure probability
\begin{equation}
P_e \ge \frac{1}{4}\,\frac{1}{1 + 4\sigma(n)^2}
\label{e15}
\end{equation}
which means that when the field is quantized it is impossible to implement the transformation (\ref{fourteen}) exactly.  Note the similarity between Eq.~(\ref{e15}) and the contribution of phase fluctuations to the error probability estimated in Eq.~(8) of \cite{jgb1}, for a state with the optimum $\sigma(\phi)^2 = 1/4\sigma(n)^2$.  This is elaborated further in Section IV below.

\subsection{Two-level atom and multimode field.}

After the publication of \cite{jgb1}, Itano \cite{itano} called into question the applicability of the results obtained from an ``effective single mode'' model to the important situation of an atom in free space, and argued that for such a system spontaneous emission would be the only source of error for quantum logic, provided the field was in a coherent state.  Although in a certain sense Itano's claim regarding spontaneous emission is accurate, it was shown in \cite{jgb2} that the constraints obtained from the Jaynes-Cummings model still apply.  This can be seen easily in the present formalism by considering the Hamiltonian
\begin{equation}
H = \hbar \sum_{\bf k} \Delta_{\bf k} a^\dagger_{\bf k}a_{\bf k} + i\hbar\sum_{\bf k} g_{\bf k}\left( a_{\bf k}\otimes\ket e \bra g - a_{\bf k}^\dagger \otimes \ket g \bra e\right)
\label{e16}
\end{equation}
for an atom interacting with many quantized field modes, labeled by the index {\bf k}, through coupling constants $g_{\bf k}$.  This can describe, in particular, an atom in free space interacting with a laser field, and allowed to emit spontaneously into the available vacuum modes. 

It is straightforward to verify that for the Hamiltonian (\ref{e16}), with the same identification $\ket e \to \ket 1$, $\ket g \to \ket 0$ as before, the following quantity is conserved
\begin{equation}
-L = -\sigma_z + 2 \sum_{\bf k} a_{\bf k}^\dagger a_{\bf k}
\label{e17}
\end{equation}
and therefore a constraint of the form (\ref{eleven}) holds, with $L_2 = -2\sum_{\bf k} n_{\bf k}$.  If the initial field state is a product state over all the modes, then $\sigma(L_2)^2 = 4\sum_{\bf k} \sigma(n_{\bf k})^2$.  If, moreover, all the modes are in a coherent state (of which the vacuum is a special case), then $\sigma(n_{\bf k})^2 = \bar n_{\bf k}$, where $\bar n_{\bf k}$ is the average number of photons in mode {\bf k}, so the constraint takes the form
\begin{equation}
1 - F^2 \ge \frac{1}{4} \,\frac{1}{1+4\bar n}
\label{e18}
\end{equation}
where $\bar n$ is the total number of photons in the field.  This can be directly compared with other multimode estimates such as \cite{jgb3} and \cite{barnes}.

In his comment \cite{itano}, Itano pointed out a seemingly unphysical aspect of the result (\ref{e18}), namely, the fact that one might apparently improve the accuracy of the operation just by making the laser beam wider, while keeping its intensity constant: this would formally increase the number of photons in the field, and hence decrease the right-hand-side of (\ref{e18}), but in fact it would not change at all the evolution of the atomic system.  This point is certainly true, but it does not mean that Eq.~(\ref{e18}) is an incorrect lower bound, only that it is not (in general) a very \emph{tight} lower bound. As shown in \cite{jgb2}, when the effects of spontaneous emission are properly taken into consideration, as Itano suggested, one finds a constraint on the fidelity that can be put in the same form as (\ref{e18}), only with a generally smaller number of photons $\bar n'$, of the order of the number of photons actually intercepted by the atom's scattering cross section.  That is to say, one typically has
\begin{equation}
1 - F^2 \ge \frac{1}{\bar n'} > \frac{1}{4} \,\frac{1}{1+4\bar n}
\label{e19}
\end{equation}

While the question of how to achieve the lower bound (\ref{e18}) will not be taken up in detail until the following Section, we may already point out that there is, clearly, a lot of information in the Hamiltonian (\ref{e16}) that does not show up anywhere in the conservation law or in the constraint (\ref{e18}): this includes, in particular, all the coupling constants $g_{\bf k}$, which account for the experimental geometry and, ultimately, for whether the coupling of the atom to the field is optimal or not.  It is natural to expect that (\ref{e18}) will only be a tight lower bound on $1-F^2$ when the coupling between the field and the atom is as close to optimal as possible.  In free space, for an atom interacting with a paraxial beam, the coupling is far from optimal and there are a lot of ``wasted'' photons; hence it is not surprising that a tighter lower limit on $1-F^2$ should be, under those circumstances, rather larger than the one obtained from the conservation law alone. Note, however, that one could, in principle, account for this suboptimal coupling in a somewhat ``ad hoc'' way by modifying the conserved quantity (\ref{e17}) so as to include only the photons in those modes that effectively couple to the atom, by introducing an appropriate ``effective mode decomposition''that is better adapted to the geometry of the atomic radiation pattern (for instance, one based on vector spherical harmonics, as opposed to plane waves). 

\subsection{Raman-coupled three-level atom.}

A popular arrangement for quantum logic involves a three-level atom in which leves $\ket{g_1}$ and $\ket{e}$ are coupled by a dipole transition (and, experimentally, by a laser field of suitable frequency and polarization), and levels $\ket{g_2}$ and $\ket{e}$ are also coupled by a dipole transition (and a laser field of their own).  The qubit states would then be $\ket{g_1}\equiv\ket 0$ and $\ket{g_2}\equiv\ket 1$, so that now
\begin{equation}
\sigma_z = \ket{g_1}\bra{g_1} - \ket{g_2}\bra{g_2}
\label{e20}
\end{equation}
The simplest Hamiltonian for this system involves two field modes, $a$ and $b$, each with a possible detuning,
\begin{align}
H = &\hbar\Delta_1 a^\dagger a + \hbar\Delta_2 b^\dagger b \notag\\
&+ i\hbar g_a \left(a\otimes\ket e \bra{g_1} - a^\dagger \otimes \ket{g_1} \bra e\right) \notag\\
&+ i\hbar g_b \left(b\otimes\ket e \bra{g_2} - b^\dagger \otimes \ket{g_2} \bra e\right)
\label{e21}
\end{align}
One has now, for the commutator $[\sigma_z,H]$,
\begin{align}
[\sigma_z,H] = &-i\hbar g_a\left(a\otimes\ket e \bra{g_1} + a^\dagger \otimes \ket{g_1} \bra e\right) \notag\\
&+ i\hbar g_b \left(b\otimes\ket e \bra{g_2} + b^\dagger \otimes \ket{g_2} \bra e\right)
\label{e22}
\end{align}
whereas
\begin{align}
[a^\dagger a,H] &= -i\hbar g_a\left(a\otimes\ket e \bra{g_1} + a^\dagger \otimes \ket{g_1} \bra e\right) \notag \\
[b^\dagger b,H] &= -i\hbar g_b\left(b\otimes\ket e \bra{g_2} + b^\dagger \otimes \ket{g_2} \bra e\right) 
\label{e23}
\end{align}
and hence it is clear that the appropriate conserved quantity is
\begin{equation}
L = \sigma_z + b^\dagger b - a^\dagger a
\label{e24}
\end{equation}
(Note that this could have been obtained from conservation of angular momentum considerations, if, for instance, the states $\ket{g_1}$ and $\ket{g_2}$ corresponded to magnetic quantum numbers $m=1$ and $m=-1$; then the photons in modes $a$ and $b$ must have opposite circular polarizations, and hence opposite angular momentum components along the quantization axis.)  Using again (\ref{eleven}) with $L=b^\dagger b - a^\dagger a$ and the assumption that the modes are initially independent, one obtains the constraint
\begin{equation}
1 - F^2 \ge \frac{1}{4} \,\frac{1}{1+\sigma(n_a)^2+\sigma(n_b)^2}
\label{e25}
\end{equation}
or, for coherent states
\begin{equation}
1 - F^2 \ge \frac{1}{4} \,\frac{1}{1+\bar n_a+\bar n_b}
\label{e26}
\end{equation}

This generalizes immediately to a multimode, free-space case, along the same lines as in the previous subsection, by introducing modes $a_{\bf k}$ that couple to the $\ket{g_1}\to\ket e$ transition, and modes $b_{\bf k}$ that couple to the $\ket{g_2}\to\ket e$ transition; the fact that $\ket{g_1}$ and $\ket{g_2}$ are orthogonal implies that one can choose the modes so as to form two disjoint sets (e.g., they would have orthogonal polarizations).  The result would again be of the form (\ref{e25}), only with a sum over modes in the denominator of the right-hand side.

\section{Saturating the CQL.}

As pointed out in the previous Section, the CQL may be well below the actual achievable error probability for many systems.  This is because many systems (i.e., many different Hamiltonians) may, in fact, satisfy the same conservation law, and the CQL derived from it must hold for the worst as well as for the best coupling possible.  The purpose of this Section is to determine what this ``best'' possible coupling is, in the sense of the one that gets as close as possible, in practice, to the error probability (\ref{eleven}).  In doing so, we shall naturally learn what are the basic physical properties of the system that are captured by the CQL (\ref{eleven}).

To begin with, as pointed out in the previous section, we shall want to have optimal coupling, meaning a single-mode (or ``single-effective-mode'') situation; hence we start by working with the Jaynes-Cummings Hamiltonian (\ref{twelve}) and asking how close it actually gets to the limit (\ref{e15}).  The question can be addressed by noticing that the full evolution operator for the Hamiltonian (\ref{twelve}) can be written, again through direct exponentiation of $H$, as
\begin{equation}
U = \begin{pmatrix} \cos gt\sqrt{a^\dagger a} & -\sin gt\sqrt{a^\dagger a}\frac{1}{\sqrt{a^\dagger a}}a^\dagger \\ \sin gt \sqrt{a a^\dagger} \frac{1}{\sqrt{aa^\dagger}}a & \cos gt\sqrt{aa^\dagger} \end{pmatrix}
\label{e27}
\end{equation}
from which follows
\begin{align}
D &=U^\dagger \sigma_z U +\sigma_x \notag\\
&= \begin{pmatrix} \cos 2gt\sqrt{a^\dagger a} & 1-\sin 2 gt\sqrt{a^\dagger a} \frac{1}{\sqrt{a^\dagger a}}a^\dagger \\ 1-\sin 2 gt\sqrt{a a^\dagger} \frac{1}{\sqrt{a a^\dagger}}a & - \cos 2gt\sqrt{a a^\dagger} \end{pmatrix}
\label{e28}
\end{align}
Now, we want the dispersion $\sigma(D)^2 = \av{D^2}-\av{D}^2$ of $D$ in a state such as $\ket{+y}\ket{\psi}$, where $\ket{+y} = (\ket 0 + i \ket 1)/\sqrt 2$ is an eigenstate of $\sigma_y$ (for which the right-hand side of (\ref{nine}) is maximized), and $\ket\psi$ is an arbitrary field state.  Actually, we shall write
\begin{equation}
\ket\psi = \sum_n C_n \ket n
\label{e29}
\end{equation}
with real coefficients $C_n$, since we got the ideal transformation (\ref{fourteen}) by letting the classical field be real. In this case we find that
\begin{equation}
\av{D} = \frac{1}{2}\left(\av{\cos 2 gt\sqrt n}-\av{\cos 2 gt\sqrt{n+1}}\right)
\label{e30}
\end{equation}
For a state such as a coherent state with a large average number of photons $\bar n$, we expect that (\ref{e30}) will be of the order of $1/\bar n$, since it involves the difference between $\sqrt n$ and $\sqrt{n+1}$, which is of the order of $\bar n^{-1/2}$, multiplied by $gt$, and we expect $gt\simeq \pi/4\sqrt{\bar n}$ at the time when the approximate transformation (\ref{fourteen}) is realized.  Then, when we square (\ref{e30}) to compute $\sigma(D)^2$, we will obtain something of the order of $1/\bar n^2$, which for large $\bar n$ is negligible versus the leading terms in (\ref{e15}) (of the order of $1/\bar n$). Hence, we will neglect this term in what follows.

For the square of $D$, we obtain, from Eq.~(\ref{e28}), the simple result
\begin{align}
D^2 = &2 - \sin 2 gt\sqrt{a^\dagger a} \frac{1}{\sqrt{a^\dagger a}}a^\dagger - \sin 2 gt\sqrt{a a^\dagger} \frac{1}{\sqrt{a a^\dagger}}a \notag\\
&+ \sigma_x  \left(\cos 2gt\sqrt{a^\dagger a} - \cos 2gt\sqrt{a a^\dagger}\right)
\label{e31}
\end{align}
and thus we have, in the state $\ket{+y}\ket{\psi}$,
\begin{equation}
\sigma(D)^2 \simeq \av{D^2} = 2 - 2\sum C_n C_{n+1}\sin(2 gt\sqrt{n+1})
\label{e32}
\end{equation}
Naturally, by the normalization of $\ket\psi$, we must have $\sum C_n^2 = 1$.  Now we may consider the various reasons why the second term on the right-hand side of (\ref{e32}) might not be exactly equal to 1 around the time $t=\pi/4g\sqrt{\bar n}$.  One reason is the spread in Rabi frequencies, $g\sqrt{n}$: for different values of $n$, it is impossible to make all the sines in the sum (\ref{e32}) equal to 1 at the same time.  Another reason is that we are adding the products $C_n C_{n+1}$, rather than $C_n^2$.  The first reason clearly means that the system is sensitive to \emph{intensity fluctuations} (uncertainty in the photon number).  The second reason, as we shall show below, captures the system's sensitivity to phase fluctuations.  

To estimate how much of $\sigma(D)^2$, in the Jaynes-Cummings model, is due to each of these two sources of error, we can write $C_n C_{n+1} = C_n^2 + (C_n C_{n+1} - C_n^2)$, and $\sin(2 gt\sqrt{n+1}) = \sin(2 gt\sqrt{\bar n+1}) + (\sin(2 gt\sqrt{n+1})-\sin(2 gt\sqrt{\bar n+1}))$, and expand (\ref{e32}), neglecting the product of terms expected to be small. The result is
\begin{align}
\sigma(D)^2 \simeq &2-2 \sin(2 gt\sqrt{\bar n+1}) \sum_n C_n C_{n+1} \notag\\
&-2 \sum_n C_n^2 (\sin(2 gt\sqrt{n+1})-\sin(2 gt\sqrt{\bar n+1}))
\label{e33}
\end{align}
Now let $2 gt\sqrt{\bar n+1} = \pi/2$.  The second sum can be approximated by writing $n = \bar n + \Delta n$, and expanding $\sin(2 gt\sqrt{n+1})$ in a power series in $\Delta n/(\bar n +1)$, to second order; the sum $\sum C_n^2 (\Delta n)^2$ then equals $\sigma(n)^2$.  The result is
\begin{equation}
\sigma(D)^2 \simeq 2 - 2\sum_n C_n C_{n+1} + \frac{\pi^2}{16 \bar n^2}\sigma(n)^2
\label{e34}
\end{equation}
The first sum, on the other hand, can be expressed in terms of the Susskind-Glogower ``exponential of phase'' operators \cite{suss}, $\exp(i\phi) = \sum \ket n\bra{n+1}$, as $\av{\exp(i\phi)}$.  Assuming that the phase of the field is a reasonably well-defined quantity with zero average, this can be expanded as
\begin{equation}
\sum_n C_n C_{n+1} = \av{\exp(i\phi)} \simeq 1-\frac{\av{\phi^2}}{2} = 1 -\frac{\sigma(\phi)^2}{2}
\label{e35}
\end{equation}
We thus get the final result 
\begin{equation}
\sigma(D)^2 \simeq \frac{\pi^2}{16 \bar n^2}\sigma(n)^2 + \sigma(\phi)^2
\label{e36}
\end{equation}
for the contribution of the field intensity and phase fluctuations to the gate error in the Jaynes-Cummings model (compare this to the result of a semiclassical calculation in Eq.~(8) of \cite{jgb1}).  Assuming the approximate validity of $\sigma(\phi)^2 \ge 1/4\sigma(n)^2$ (which, for a coherent state with large $\bar n$, holds as an approximate equality), we get
\begin{equation}
\sigma(D)^2 \ge \frac{\pi^2}{16 \bar n^2}\sigma(n)^2 + \frac{1}{4\sigma(n)^2}
\label{e36a}
\end{equation}
and it is clear that the second term in (\ref{e36a}) is very much like the conservation-law constraint (\ref{ten}), with $L_2 = (\pm)2n$, as long as one can assume $\sigma(n)^2 \gg 1$.  Again, this is automatically the case for a large-$\bar n$ coherent state, and, at any rate, it is the most interesting case, since one typically will want the QCL to be very small, and this can only happen if $\sigma(n)^2 \gg 1$.   

On the other hand, there is clearly nothing like the first term in (\ref{e36}) in the conservation-law constraint (\ref{ten}); that is, there is no indication of an error that \emph{grows} with the size of the intensity fluctuations.  This is simply because the atom-field coupling that results from the Hamiltonian (\ref{twelve}) has an intrinsic photon-number dependence that is \emph{not} a necessary consequence of the conservation law (\ref{one}).  One can, in fact, envision a whole family of Hamiltonians compatible with the same conservation law, of the general form
\begin{equation}
H = i\hbar g \sum_n\left(f_n \ket n \bra{n+1} \otimes \ket e \bra g - f_n \ket{n+1} \bra n \otimes \ket g \bra e \right)
\label{e37}
\end{equation}
where the $f_n$ are arbitrary numbers.  The Jaynes-Cummings model corresponds to the choice $f_n = \sqrt{n+1}$, but one could equally well (formally, at least) remove all dependence on the photon number by setting $f_n =1$ for all $n$.  Recalling again the Susskind-Glogower exponential-of-phase operator, we see that the resulting Hamiltonian is just what one would obtain by quantizing \emph{only the field phase} in the semiclassical Hamiltonian $
H = i\hbar g E_0 (e^{i\phi} \ket e \bra g - e^{-i\phi}  \ket g \bra e )$ discussed in connection with Eq.~(\ref{thirteen}):
\begin{equation}
H = i\hbar g \sum_n\left(\ket n \bra{n+1} \otimes \ket e \bra g - \ket{n+1} \bra n \otimes \ket g \bra e \right)
\label{e38}
\end{equation}
We could continue the discussion with the Hamiltonian (\ref{e38}) (see Appendix B for details), but it is more transparent at this point to work with a hypothetical Hermitian phase operator that satisfies
\begin{equation}
[\hat \phi,\hat n] = -i
\label{e40}
\end{equation}
and rewrite our (also hypothetical) interaction Hamiltonian (\ref{e38}) as
\begin{equation}
H = i\hbar g \left(e^{i\hat\phi} \ket e \bra g - e^{-i\hat\phi}\ket g \bra e \right)
\label{e41}
\end{equation}
(for clarity, in the remainder of this section the operators $\hat \phi$ and $\hat n$ have been given a carat). 

Apart from making the correspondence with the semiclassical case more obvious, the form (\ref{e41}) has the satisfactory feature of making the conservation law (\ref{one}) straightforward, since by (\ref{e40}) the operators $e^{\pm i\hat\phi}$ are ``displacement operators'' for the photon number $\hat n$, changing it by one unit up or down.  Thus, in a sense, Eq.~(\ref{e41}) is the simplest nontrivial Hamiltonian compatible with a conservation law of the form (\ref{one}): it just captures the essential fact that the photon number must change by $\pm 1$ as the atom makes a transition up or down.  

It is known (again, see \cite{barnett} for details) that there are difficulties with the postulated Hermitian operator $\hat\phi$, essentially, because one cannot lower the number of photons below the vacuum level ``$\ket 0$.'' Pegg and Barnett have defined a Hermitian $\hat\phi$ for which something like (\ref{e40}) approximately holds for most relevant, ``physical'' states \cite{barnett2}.  One of us \cite{ozawaphase} has also shown that it is possible to define a Hermitian phase operator in a larger Hilbert space, augmented by ``macroscopic'' states with a (formally) infinite photon number, using nonstandard analysis.  We shall not go into these complications here, but merely assume that something like Eqs.~(\ref{e40}) and (\ref{e41}) makes sense, in an appropriate limit, for useful, physical states (as the derivation in Appendix B shows, it is enough to assume that the contribution of the vacuum state to the field is negligible).

The evolution operator associated to the interaction Hamiltonian (\ref{e41}) is then just like the semiclassical result (\ref{thirteen}), only with a quantized $\hat \phi$:
\begin{equation}
U(t) = \cos(g t) + \begin{pmatrix} 0 & -e^{-i\hat\phi} \\ e^{i\hat\phi} & 0 \end{pmatrix} \sin(g t)
\label{e42}
\end{equation}
and thus at the time $t=\pi/4g$ it becomes
\begin{equation}
U = \frac{1}{\sqrt 2}\begin{pmatrix} 1 & -e^{-i\hat\phi} \\ e^{i\hat\phi} & 1 \end{pmatrix} 
\label{e43}
\end{equation}
In an initial state $\ket{+y}\ket\psi$, the difference operator $D = U^\dagger \sigma_z U + \sigma_x$ has the expectation value $\av D = \av{\sin\hat\phi}$, which we can set equal to zero, and then we find
\begin{equation}
\sigma(D)^2 = \av{D^2} = 2 - 2\av{\cos\hat\phi}
\label{e44}
\end{equation}
To minimize $\sigma(D)^2$, we clearly want the phase distribution for the field state $\ket\psi$ to be centered at zero and as narrow as possible.  Under this assumption of small phase dispersion (which clearly implies large $\sigma(\hat n)^2$, by (\ref{e40})), we can expand (\ref{e44}) and get
\begin{align}
\sigma(D)^2 &\simeq   2 - 2\left(1-\frac{\av{\hat\phi^2}}{2}\right) \notag \\
&=\sigma(\hat\phi)^2 \notag \\
&\ge \frac{1}{4\sigma(\hat n)^2} 
\label{e45}
\end{align}

We have, therefore, finally succeeded at devising a situation where the lower limit $\sigma(D)^2 \ge 1/(1+4\sigma(n)^2)$ can be approximately reached (assuming large $\sigma(n)^2$).  The necessary ingredients were:  (1) an effective single-mode interaction; (2) an atom-field coupling that is insensitive to intensity (i.e., photon number) fluctuations, and (3) a state with minimum uncertainty in the phase-intensity variables.  In the large $\bar n$ (and small $\sigma(\phi)^2$) limit, the CQL clearly captures the error due to (quantum) field \emph{phase} noise. 

\section{Conclusions}

In this paper, we have presented a number of systems, involving the interaction of an atom with the quantized electromagnetic field, for which constraints on the accuracy of quantum logical operations can be derived from conservation laws of the form first discussed by one of us in \cite{ozawa1}.  These constraints would appear to include those arising from atom-field entanglement (although we have not explored this isssue in detail), and, in the appropriate limit, they seem to include as well those arising from fluctuations in the field phase.  The special significance of the phase is clear from the ``minimal Hamiltonian'' (\ref{e41}) compatible with the conservation law (\ref{one}):  as the conjugate variable to the photon number, which is the field quantity that appears in the conservation law, it must be involved in the Hamiltonian in the way given by (\ref{e41}), to ensure that photons are created and destroyed appropriately as the atom makes a transition.  This leads to its unavoidable presence in the evolution operator, as in Eq.~(\ref{e43}), which then naturally becomes sensitive to phase fluctuations. 

It follows from the above that quantum logical operations that are \emph{not} constrained by a CQL of the form considered here must be (at least to lowest order) \emph{insensitive to fluctuations in the field phase}.  

On the other hand, we \emph{cannot} conclude that operations not constrained by the CQL can be performed without any errors, since the possibility still remains that they might be sensitive to \emph{intensity} fluctuations, whose effects, as shown in the previous section, are not accounted for by the CQL.  Whether this is the case or not, and if so, to what extent, clearly needs to be ascertained separately for each specific physical system under consideration; the present work merely provides a starting point by showing why such an undertaking would be necessary. (See, in this context, the discussion in \cite{lidar,ozawa3}.)

Finally, it is not yet, at this point, entirely clear how these considerations may relate to the ultimate limits to ``quantum logic with quantum controls'' discussed by one of us in \cite{jgb3}.  As elaborated in later work, it seemed that an important ingredient in the derivation of such limits was the requirement that the interaction should be ``turned on and off'' by the quantized control system itself.  It would be interesting to see how much of that work might be clarified by the present formalism; for instance, by formulating some of the constraints in terms of other conservation laws, by searching for approximate conservation laws, or by separating the constraints that follow from such laws from those that have a different origin.  We expect to be able to present results on this program in the near future.

The authors acknowledge support from the U. S. Army Research Office, the Strategic Information and Communications R\&D Promotion Scheme of the MPHPT of Japan, the CREST project of the JST, and the Grant-in-Aid for Scientific Research of the JSPS.

\appendix
\section{Derivation of the fidelity equation (\ref{eleven})}

The purpose of this Appendix is to show in detail how one can go from the noise $\sigma(D)^2$ of Eq.~(\ref{ten}) to the ``error probability'' $1-F^2$ of Eq.~(\ref{eleven}).  The derivation closely follows \cite{ozawa2}, {\em circa} Eq.~(88).  First, the action of the evolution operator $U$ in the computational basis can generally be written as 
\begin{align}
U\ket 0\ket\psi &= \ket 0\ket{E_0^0} + \ket 1\ket{E_1^0} \notag \\
U\ket 1\ket\psi &= \ket 0\ket{E_0^1} + \ket 1\ket{E_1^1}
\label{a1}
\end{align}
where the field states $\ket{E_b^a}$ are not necessarily normalized.  Next, assuming that the initial qubit state is $\ket{+y}$ (as we did in the derivation of Eq.~(\ref{ten})) we have 
\begin{align}
\av{D^2} &= \bra{\psi}\bra{+y}(U^\dagger\sigma_z U - \sigma_x)^2 \ket{+y}\ket\psi \notag \\
&= ||(U^\dagger\sigma_z U - \sigma_x)\ket{+y}\ket\psi||^2 \notag \\
&= ||(\sigma_z U - U\sigma_x)\ket{+y}\ket\psi||^2 \notag \\
&= ||\ket{E_0^0} -\ket{E_0^1}||^2 + ||\ket{E_1^0} + \ket{E_1^1}||^2
\label{a2}
\end{align}
where the last equality follows from the direct application of (\ref{a1}) to the states $\ket{+y}\ket\psi$ and $\sigma_x\ket{+y}\ket\psi$.  

Now, using again (\ref{a1}), observe that the first term of (\ref{a2}) can be written as 
\begin{equation}
||\ket{E_0^0} -\ket{E_0^1}||^2 = 2||\bra 0 U \ket{-x}\ket\psi||^2
\label{a2b}
\end{equation}
where $\ket{-x} = (\ket 0 -\ket 1)/\sqrt 2$ is the $-1$ eigenstate of $\sigma_x$; hence this term expresses the probability that the qubit may be found in state $\ket 0$ at the end of the transformation, if the initial state was $\ket{-x}$.  Note, however, that a pure Hadamard gate should turn $\ket{-x}$ into $\ket 1$ exactly; hence this term is nothing other than twice the error probability (or ``infidelity''), $1-F^2(\ket{-x})$, for an initial state $\ket{-x}$.  Similarly it can be seen that the other term in (\ref{a2}) equals $2(1-F^2(\ket{+x}))$.  We conclude that
\begin{equation}
\av{D^2} = 4 - 2F^2(\ket{-x}) - 2F^2(\ket{+x}) \le 4(1-F_{min}^2)
\label{a3}
\end{equation}
when the minimum fidelity is taken over all the possible initial states of the qubit.  (Note that there is a typographical error in the corresponding equations (88) and (90) in \cite{ozawa2}: both fidelity terms must subtract from 1, as in Eq.~(\ref{a3}) above.)

Finally, noting that $\av{D^2} \ge \sigma(D)^2$ always holds, we obtain
\begin{equation}
1-F_{min}^2 \ge \frac{1}{4}\av{D^2} \ge \frac{1}{4}\sigma(D)^2
\label{a4}
\end{equation}
which is Eq.~(\ref{eleven}).

\section{Derivation of (\ref{e45}) without assuming a Hermitian phase operator}
The evolution Hamiltonian (\ref{e38}) can be easily integrated to yield the evolution operator
\begin{align}
U=&\cos gt \notag\\
&+ \begin{pmatrix} 2 \sin^2(gt/2) \ket 0\bra 0  & -\sin gt \sum_n\ket{n+1}\bra n \\ \sin gt \sum_n \ket n \bra{n+1} & 0 \end{pmatrix}
\label{b1}
\end{align}
In the expression above, the state $\ket 0$ represents the photon vacuum.  The effect of this extra term can be made negligible if one works with a field state that has a negligible projection on the vacuum; for instance, for a coherent state, $C_0 = e^{-\bar n/2}$.  Neglecting it altogether, then, we see that $\sigma_z$ is transformed as
\begin{align}
U^\dagger \sigma_z &U = \notag\\
&\begin{pmatrix} \cos 2 gt  & -\sin 2gt \sum_n\ket{n+1}\bra n \\ -\sin 2gt \sum_n \ket n \bra{n+1} & -\cos 2gt \end{pmatrix}
\label{b2}
\end{align}
Without any further approximations, one then obtains, for the expectation value of $D=U^\dagger \sigma_z U+\sigma_x$, in the state $\ket{+y}\ket\psi$,
\begin{equation}
\av{D} = -i\sin(2 gt)\left\langle\sum_n\bigl(\ket{n+1}\bra n -\ket n\bra{n+1}\bigr)\right\rangle
\label{b3}
\end{equation}
which equals zero if the state $\ket\psi =\sum_nC_n\ket n$ is chosen to have all real coefficients $C_n$ (zero average phase).  For the operator $D^2$, one gets
\begin{align}
D^2 = &2 -\sin(2 gt) \sum_n\bigl(\ket{n+1}\bra n +\ket n\bra{n+1}\bigr) \notag\\
&-\sin^2(2 gt)\ket 0\bra 0 \otimes\begin{pmatrix} 1 & 0 \\ 0 & 0 \end{pmatrix}  
\label{b4}
\end{align}
Thus, neglecting again the field's vacuum-state contribution, we get for $\sigma(D)^2$
\begin{equation}
\sigma(D)^2 = 2 - 2\sum C_n C_{n+1}\sin(2 gt) = \sigma(\phi)^2 \ge \frac{1}{4\sigma(n)^2}
\label{b5}
\end{equation}
at the time $gt = \pi/4$, where the connection with phase fluctuations is established via the Susskind-Glogower operators, as in the derivation of (\ref{e35}).

\end{document}